# Light Reconfigurable Geometric Phase Optical Element with Multi-stable States

Revised 10/17/2018 22:29:00

Xiao-Qian Wang[1], Alwin Ming-Wai Tam[1,2], Wei-Qiang Yang[1], Engle Liao[2], Ka Chun Cheung[2], Wei Hu[3], Dong Shen[1], Vladimir Chigrinov[2], Hoi-Sing Kwok[2], Zhi-gang Zheng[1,*], Yanqing Lu[3,†], and Quan Li[4,#]

[1] *Department of Physics, East China University of Science and Technology, Shanghai 200237, China*
[2] *State Key Laboratory on Advanced Displays and Optoelectronics Technologies, Hong Kong University of Science and Technology, Hong Kong, China*
[3] *National Laboratory of Solid State Microstructures, Collaborative Innovation Center of Advanced Microstructures and College of Engineering and Applied Science, Nanjing University, Nanjing 210093, China*
[4] *Liquid Crystal Institute and Chemical Physics Interdisciplinary Program, Kent State University, Kent, Ohio 44242, United States*

We present the design methodology of a light reconfigurable geometric phase optical element with multi-stable diffraction efficiency states, enabled by a photoresponsive self-organized chiral liquid crystal. Experimental demonstration shows the device exhibits a broad diffraction efficiency tunable range that can be smoothly modulated under alternate stimulation of ultraviolet and green lights. Distinctive to previous designs, the regulation of diffraction efficiency fundamentally stems from the modulation of geometric phase together with dynamical phase retardation, and any intermediate diffractive state is memorized. Such multi-stability facilitates applications including energy-saving all-optical signal processing in classical and quantum level, and phase hologram for anti-counterfeit.

In polarization optics, it is notable that light propagating through an optical anisotropic material will lead to changes in polarization state. This phenomenon is well described diagrammatically on the Poincaré sphere, for which the change in polarization state traces out a path trajectory on the sphere [1]. As introduced by Pancharatnam and Berry over two decades ago, changes in the state of polarization of light will in general induce a geometric phase (GP) (also known as the Pancharatnam-Berry phase) [2,3]. A classic geometric relation of the phase is when the evolution of the polarization state sketches out a closed loop on the Poincaré sphere, the geometric phase induced is given by half of the solid angle subtended by the enclosed path [2,3]. The phase is distinctive from the conventional dynamical phase arising from optical path.

As first introduced by Bhandari [4], geometric phase optical elements (GPOE) that uses GP to modulate the wavefront of the incident beam is a flat diffractive optical element featured by space variant dielectric or absorptive anisotropic structure on the incident surface. Since then, many experimental demonstration of GPOE using subwavelength gratings [5], metasurfaces based on metallic and dielectric nanoscatterers [6-9], liquid crystal (LC) devices with inhomogeneous alignment [10-19] have been reported. GPOE possesses a unique property for which the GP profile derived from the distribution of spatially variant optics axis, is dependent on the handedness of circular polarization (CP) or photon spin angular momentum (SAM). The specific polarization dependent property of GPOE plays a vital role in the field of classical and quantum optics including the generation of polarization dependent vortex beam for quantum information processing [20], demonstration of polarization sensitive hologram [21], and optical phenomenon discovery [16]. Among the aforementioned materials, optical states associated with various diffraction efficiencies (DE) can only be enabled by LC based GPOE using external electric field [22-24]. LC-GPOE with broad DE modulation range enables switching between diffractive 'off' and 'on' states opening various 'smart' optical signal processing applications [23,24].

More recently, chiral-LC (CLC) GPOE has been introduced in Ref. [25-27], for which the device comprises of spatially variant optics axis rotated in the perpendicular direction with respect to the device plane, forming a helical structure. When the wavelength of incident light $\lambda$ is within the Bragg regime of the helix with pitch $P$ (rotation of LC director by $2\pi$) i.e $\lambda \sim P$, the device behaves as a CP selective reflective GPOE [25,26]. While beyond the Bragg regime ($\lambda \ll P$), the GPOE is a transmissive type and with the use of double twisted chiral layers, the device possesses high DE over a broad bandwidth for a designed pitch [27]. Despite of the aforementioned advantages of LC and CLC GPOE, these devices require an external field to sustain their state of operation, leading to high power consumption and susceptibility to electrical fault. Therefore, there is an urgent demand for LC GPOE possessing multi-stable states to address these issues. Recent studies in Ref. [28,29] has shown CLC with photoresponsive helical pitch exhibits multi-stable configuration, implying the optical response of the material is preserved when the light stimulus is terminated. Thus, photoresponsive CLC is a good candidate for the realization of multi-stable LC GPOE using all-optical control.

Herein, we disclose a method and present an analytical study for the design of light reconfigurable CLC GPOE (LR-CLC-GPOE) with multi-stable diffractive states enabled by photoresponsive helical pitch operating beyond the Bragg regime. Distinctive from conventional LC-GPOE for which the GP profile is solely determined by a specific designed LC orientation in a two-dimensional plane, the proposed device utilizes the reconfiguration of GP by manipulating in a third dimension, helical twist of LC with light stimulus. To characterize the optical response for LR-CLC-GPOE, we have established a rigorous analytical model formulating a closed-form solution to determine the DE of the device. The validity of this peculiar analytical model and the unique stability of this new class of GPOE have been confirmed from the experimental demonstration on the various examples of LR-CLC-GPOE including polarization grating (PG) and geometric phase lens (GPL).

The physics of the LR-CLC-GPOE is elucidated by determining the orientation of the LC director of the device. The LC director $n$ dictates the local average orientation of LC molecules, and can be represented in the vector form $n(\theta,\phi) = \cos\theta\cos\phi\, x + \cos\theta\sin\phi\, y + \sin\theta\, z$ [30], where $\theta$ and $\phi$ corresponds to the polar angle and azimuthal angle of the director, respectively. The photo-responsive CLC used for the LR-CLC-GPOE is prepared by mixing a commercial nematic LC TEB300 (99 $wt\%$, from Slichem, China) with a homemade right-handed chiral molecular switch QL55 (molecular structure shown in Fig. 1(a), 1 $wt\%$) exhibiting two distinct isomers upon different light irradiations, *i.e.* ultraviolet (UV) and visible (VIS) lights (532 $nm$). Fig. 1(a) shows under UV irradiation, *trans*-isomer state of QL55 transforms into *cis*-isomer state, resulting in a gradual elongation of the helical pitch, while the process is reversible under illumination of green light. Under a controllable exposure dosage of UV and VIS lights, the overall helical pitch of the photo-responsive CLC mixture can be modulated within a considerable range $8.3\mu m \leq P \leq 22.1\mu m$. In the absence of light stimulus, the arrangement of CLC mixture is almost unchanged in a relatively long time due to the low concentration of photoresponsive QL55, establishing the fundamental basis of the multi-stable property of the device.

Anticipating the planar alignment on both surfaces will lead to discrete jump of the helical pitch due to strong azimuthal (planar) surface anchoring strength as in Ref. [28,29], resulting in the sharp state transition of the optical response, hybrid aligned (HA) boundary condition (BC) is adopted in the design to allow for gradual change of the helical pitch, indicating a continuous and smooth DE modulation of GPOE. The top surface of the design prototype at $z=d$ is deposited with homeotropic aligned polyimide (PI-5661), while the bottom surface at $z=0$ comprises of a photo-stable spatial variant alignment layer using sulfonic azo-dye passivated with reactive mesogens (RM). Herein, $z$ is the direction along the cell gap $d$. The polar and azimuthal anchoring strengths for both alignment materials can be regarded as 'strong' *i.e.* $>10^{-4}J/m^2$ [31,32], thus the LC molecules on the $xy$ surface yields the following BCs $\theta(z=0)=0$, $\phi(x,y,z=0)=\alpha(x,y)$, $\theta(z=d)=\frac{\pi}{2}$, $\phi_z|_{z=d} = \frac{K_{22}}{K_{33}}q_0$ (see Supplementary Material (SM) Sec. 1 for derivation [33]). Under strong surface anchoring condition, according to Frank-Oseen elastic theory, the free energy $f$ of the LC bulk can be expressed as [34]:

$$f = \frac{1}{2}K_{11}(\nabla \cdot n)^2 + \frac{1}{2}K_{22}(-n\cdot\nabla\times n + q_0)^2 + \frac{1}{2}K_{33}(-n\times\nabla\times n)^2 \quad (1)$$

where $q_0 = 2\pi/P$ is the helical wavenumber, and the elastic constants $K_{11}=4.6 pN$, $K_{22}=4.1 pN$, and $K_{33}=20.1 pN$ corresponds to splay, twist and bend elastic deformations of the photo-responsive CLC mixture, respectively (See SM in Sec. 2A for details measured parameters).

The orientation of the LC director $n$, determined by the minimization of free energy $f$ of the system is calculated using calculus of variation, which gives the analytical expression for the orientation angles $\theta$ and $\phi$ (refer to SM Sec. 1A for derivations [33]),

$$\theta_z = \frac{1}{\sqrt{K_{11}\cos^2\theta + K_{33}\sin^2\theta}}\sqrt{Q - \frac{q_0^2 K_{22}^2 \cos^2\theta}{K_{22}\cos^2\theta + K_{33}\sin^2\theta}} \quad (2)$$

$$\phi(x,y,z) = \alpha(x,y) + \Phi(z) \quad (3a)$$

$$\Phi_z = \frac{K_{22}q_0}{K_{22}\cos^2\theta + K_{33}\sin^2\theta} \quad (3b)$$

Here, $\Phi$ is the twist angle of the helical system, $\alpha(x,y)$ is the spatial alignment orientation, and $Q$ in Eq. (2) is a constant (determine from BCs). The letter subscript '$z$', denotes the partial derivatives with respect to $z$. The expressions in Eq. (2 and 3) shows the tilt angle $\theta$ and twist angle $\Phi$ are independent on the $x, y$ directions, which is also illustrated in Fig. 1(b). This is indeed a simplification of the situation as the spatial alignment coupling will in fact lead to variation of $\theta$ and $\Phi$ in the $xy$ plane. Though, the simplified expressions in Eq. (2 and 3) are valid provided the helical pitch $P$ and the cell gap $d$ are several times smaller than the minimum local pitch $\Lambda_L$ of the alignment layer *i.e.* $d, P \ll \min(\Lambda_L)$ (See SM [33] Sec. 1A for detail derivation). We define this situation as the adiabatic alignment variation (AAV) condition. Nevertheless, the nonlinear coupled differential equations for angles $\theta$ and $\Phi$ in Eq. (2 and 3) satisfying the specified BCs must be solved numerically.

The optical response of LR-CLC-GPOE can be determined from the numerical solutions of $\theta$ and $\phi$ in Eq. (2 and 3). Since the relation between the incident wavelength and the pitch of the CLC mixture is far from the Bragg reflection regime, the device can be determined using Jones calculus that divides the thickness of the LC bulk into $N$ sublayers. After some algebraic manipulations, the transmitted output wave $E_t$ is mathematically described by the general form:

$$E_t = \vec{T}(x,y) E_{in}$$
$$= \left\{ B_N \left(-i\vec{\sigma_1} + \vec{\sigma_3}\right) \exp\left[i2(\alpha(x,y)+\Phi_N)\right] \right.$$
$$\left. + C_N \left(i\vec{\sigma_1} + \vec{\sigma_3}\right) \exp\left[-i2(\alpha(x,y)+\Phi_N)\right] + \left(A_N \vec{I} + D_N \vec{\sigma_2}\right) \right\} E_{in} \quad (4)$$

Here, $E_{in}$ is the polarization state of the incident light, $\vec{T}$ is the transmission matrix that is constructed by the Pauli spin matrices $\vec{\sigma_1} = \begin{bmatrix} 0 & 1 \\ 1 & 0 \end{bmatrix}$, $\vec{\sigma_2} = \begin{bmatrix} 0 & -i \\ i & 0 \end{bmatrix}$ and $\vec{\sigma_3} = \begin{bmatrix} 1 & 0 \\ 0 & -1 \end{bmatrix}$, $\vec{I}$ is the 2×2 identity matrix. $A_N$, $B_N$, $C_N$, $D_N$ are the iterative coefficients, dependent on angles $\theta$, $\Phi$ and retardation $\Gamma$ (see Sec. 1B of SM for detail derivation [33]). Herein, the subscript integer $N$ represents the last sublayer of the GPOE. The angle $\Phi_N$ is the total twist angle of the helical system. Note that the first/second term in Eq. (4) consists of a positive/negative complex exponential i.e. $\exp[\pm i2(\alpha(x,y)+\Phi_N)]$, which corresponds to GP associated with plus/minus first ($\pm 1^{st}$) order diffracted beam for left/right handed CP (LHCP/RHCP) incident wave or +/- photon SAM, respectively. The remaining last two terms in Eq. (4) corresponds to the zeroth ($0^{th}$) order undiffracted transmitted beam (no propagation deviation). Eq. (4) indicates the total GP comprises of an invariable component determined by the alignment profile $\alpha(x,y)$, and a variable component, twisted angle ($\Phi$) of CLC, induced by the change of helical pitch ($P$) upon light stimulus, which leads to a light reconfigurable GP. Even though $\Phi$ changes with $P$, it is always approximately constant across the $xy$ plane, thus the propagation direction of $\pm 1^{st}$ order is primarily governed by $\alpha(x,y)$. Nevertheless, the light modulation of CLC helix, which leads to changes in helical wavenumber $q_0$ in Eq. (2 and 3b), results in the variation of the GP (twist angle $\Phi$) and dynamical phase retardation (polar angle $\theta$), and thus modifies $A_N$, $B_N$, $C_N$, $D_N$ and consequently $\pm 1^{st}$ order DE is regulated. Since $C_N = B_N^*$ (refer to Eq. S39b and c in SM [33]), either measurement of $\pm 1^{st}$ order DE for corresponding CP incident wave yields the same result, and can be expressed as,

$$\eta_{\pm 1} = 1 - E_{in}^* \left(A_N^* \vec{I} + D_N^* \vec{\sigma_2}^*\right)\left(A_N \vec{I} + D_N \vec{\sigma_2}\right) E_{in} \quad (5)$$

Hence, the analytical model highlights the feasibility to reconfigure DE of LR-CLC-GPOE under light stimulus, as shown in Fig. 1(c), from the underlying physics of simultaneous modulation of GP and dynamical phase retardation. Moreover, although the eigenbasis of $0^{th}$ order transmitted beam is CP, the helicity of the device fundamentally manifests a $\Phi$ dependent phase difference between the $0^{th}$ order beam from LHCP and RHCP incident wave or +/- photon SAM. Such phase difference is the source for non-vanishing $\vec{\sigma_2}$ related term (last term) in Eq. (4), implying the polarization state of incident wave and the $0^{th}$ order beam are in general different. Thus, the proposed device is fundamentally distinctive from conventional LC-GPOE with unmodulated $0^{th}$ order transmitted beam, and dynamical phase retardation modulated DE under applied field [22-24].

It is worth mentioning the iterative approach in Eq. (4) is universal, and will converge to the specific analytic form as for the case of conventional LC-GPOE [15], and CLC-GPOE with planar alignment [27] when the corresponding angles $\theta$, $\Phi$ are used. The model is valid provided the AAV condition holds, and the incident wavelength $\lambda$ is beyond Bragg regime.

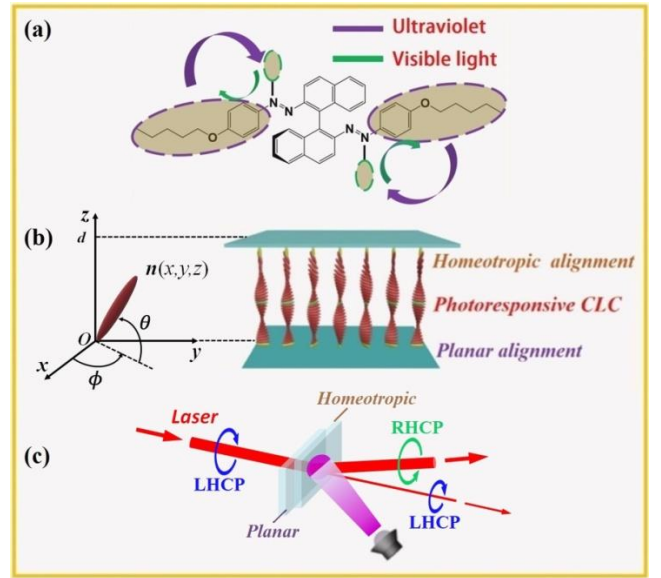

FIG. 1. (Color online) (a) Molecular structure of azobenzene-based binaphthyl chiral switch QL55 doped into the photo-sensitive CLC. Under UV irradiation, the 'trans' state is transformed to the 'cis' state resulting in an elongation of helical pitch. The process is reversible by VIS irradiation at 532 $nm$. (b) LC director orientation governed by HA BCs with homeotropic alignment layer at $z = d$, and spatially inhomogeneous planar alignment layer at $z = 0$. (c) An LHCP incident beam passing through the LR-CLC-GPOE results in a $1^{st}$ order diffracted beam where the polarization state is reversed and the propagation direction is altered, as well as an undiffracted $0^{th}$ order wave. The energy distribution between $0^{th}$ and $1^{st}$ order can be controlled from VIS and UV irradiations.

To verify the proposed analytical design of the LR-CLC-GPOE, two types of GPOE, PG and GPL are fabricated. The alignment distribution $\alpha(x,y)$ is given by $\alpha_{PG}(x) = \frac{2\pi}{\Lambda} x$ and $\alpha_{GPL}(x,y) = \frac{\pi}{2f\lambda}(x^2+y^2)$ respectively, where $x$ and $y$ represents two orthogonal directions with respect to the device plane. The grating pitch of the fabricated PG prototype is $\Lambda = 125\,\mu m$, while the focal distance of the GPL prototype is $f = 200\,mm$. The thicknesses $d$ of the CLC layers for both devices are 5 $\mu m$. Indeed, the above design parameters are selected such that aforementioned adiabatic spatial alignment condition is valid. The spatial variant alignment is prepared using the GP hologram optical setup specified in Ref. [15,18] (Fabrication processes are detailed in the SM Sec. 2B [33]).

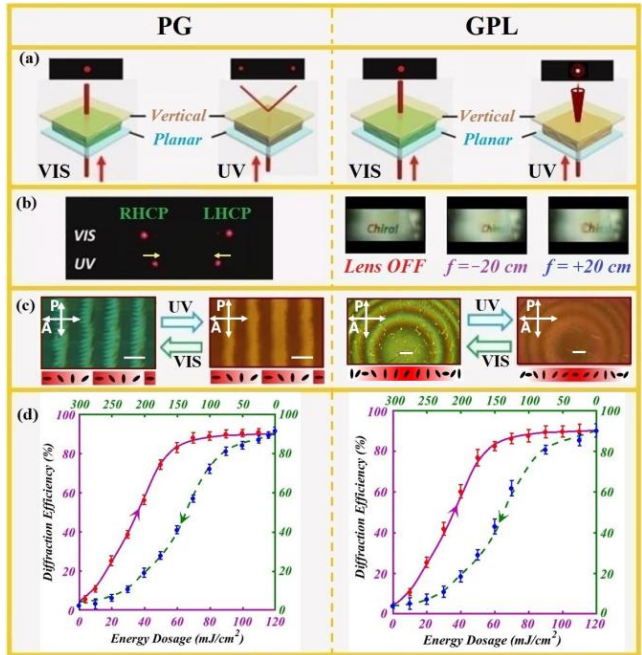

FIG. 2. (Color online) Optical characterization of the fabricated LR-CLC PG (left) and GPL (right). (a) The diffraction patterns of PG and GPL under VIS and UV irradiations for non-polarized monochromatic light at 632.8 *nm*. (b) Observation of CP dependence of the devices. Under the insertion of left/right handed circular (LHC/RHC) polarizer with UV exposure, the beam is steered to the left/right for PG, while the image is magnified/demagnified by the +/- 20 *cm* focal length. The diffraction is suppressed and the device is effectively 'off' under VIS exposure. (c) Micrographs of the LR-CLC PG and GPL captured under crossed polarizer (P) and analyzer (A) upon VIS and UV irradiations. The scale bars for PG and GPL represent 25 *μm* and 150 *μm*, respectively. The corresponding alignment orientations are indicated below the micrographs. (d) Optical modulation of DE states of LR-CLC PG and GPL with corresponding exposure UV and VIS dosages. Upon VIS/UV irradiations, the blue/red dots with corresponding error bars represent the respective measured data, while the green dashed/purple solid lines are the respective numerically calculated results.

Fig. 2(a) shows under sufficient UV light dosage, the pitch (*P*) of the helix is elongated (small total twist angle), and both PG and GPL are in high DE states, where the typical dual beam steering characteristics of PG and the bifocality of GPL subjected to a non-polarized monochromatic incident probe beam at 632.8 *nm* are confirmed [18]. Indeed, the probe wavelength falls outside the absorption spectral window of the photoresponsive chiral molecular switch QL55, and thus the helical pitch *P* is almost invariable. The CP or SAM dependence for the devices is verified in Fig. 2(b), for which the PG diffracts the probe beam to the left/right at an angle of 0.58° as expected from Bragg's law, when a left/right handed circular polarizer (LCP/RCP) is inserted in front of the device, while the letters are magnified/demagnified from the GPL when observed through the LCP/RCP substantiating the focused/defocused property of the lens (refer to SM Sec. 3D [33] for detailed experimental setup). Conversely, under sufficient VIS light dosage, the pitch is contracted, and Fig. 2(a) and (b) show the diffraction is suppressed giving rise to the diffraction 'off' state. The micrographs in Fig. 2(c), captured under crossed polarizers, shows the typical polarization interference patterns for PG and GPL as depicted from the spatial alignment profile. Upon VIS irradiation, a greenish birefringent color is observed from PG and GPL micrographs, implying the devices have low DE near 632.8 *nm* (can be verified from Eq. (4 and 5)). Moreover, fingerprint textures can be observed from the VIS exposed micrograph (refer to SM Sec. 3A for details [33]). When the devices are subsequently exposed with UV light, the fingerprint texture gradually disappears and a reddish birefringent color is observed, signifying high DE in the red wavelength regime. As a rule of thumb, high DE state $>85\%$ is established when the light reconfigurable GP, denoting the total twist angle $\Phi_N$, is less than 50 degrees and the retardation is approximately half wave, while low DE state ($<15\%$) is achieved when $\Phi_N$ is greater than 100 degrees. Upon the stimulation of UV ($\lambda_{UV} \approx 365$ *nm* at 2 *mW/cm²*) light, the red dots with corresponding error bars in Fig. 2(d) show the measured DEs of PG and GPL smoothly increase over the modulation range of 2.0 % - 92.5 % and 4.0 % - 90.2 %, respectively (refer to SM Sec. 3B [33] for detailed measurement setup). After maximum DE states are reached, the DEs decrease to the minimum DE states upon subsequent VIS ($\lambda_{VIS} = 532\pm10$ *nm* at 5 *mW/cm²*) irradiation as indicated from the blue dots with corresponding error bars. The simulated results are obtained by considering the helical pitches in Eq. (1) upon the corresponding energy dosages and subsequently calculated by the above-mentioned numerical analysis. This is in good agreement with the experimental results as shown by the green dashed (VIS) and the purple solid (UV) lines in Fig. 2(d). Note that, in the low DE regime, the measured data are slightly lower than the numerically calculated result which could be attributed to the presence of the fingerprint texture that leads to diffractive noises. Moreover, the hysteresis loop enclosed by UV and VIS DE curves of each device illustrated in Fig. 2(d) indicates the thermal dissipation during the alternating modulation process.

Indeed the DE of the device is memorized upon the termination of the light stimulus, for which the DE changes by less than 5% over 4 hours along the smooth curve in Fig. 2(d) (device stability are detailed in SM Sec. 3C). Therefore, the device can be regarded as stable over a considerable time frame. The salient multi-stable feature of the device is of great value for establishing new research development in the field of all optical polarization processing, for examples, provide additional covert security feature to anti-counterfeit optical hologram, and realizing reconfigurable cross-connects in a communication link that is resilient to sudden power failure.

In summary, LR-CLC-GPOE with smooth reconfiguration of multi-stable DE states enabled by photo-responsive CLC mixture has been proposed. A rigorous theoretical analysis characterizing the optical characteristics of the proposed device from LC director distribution presents the DE modulation is fundamentally due to the simultaneous light reconfigurable GP (twist angle) and the dynamical phase retardation. To the feasibility of the analytical design, PG and GPL have been fabricated and the results measured are in good agreement with the features depicted in the analytical result, for which broad reconfigurable DE range is achieved upon UV and VIS stimulation. It is believed the proposed device paves the way towards 'smart' optics that exhilarates research development in the field of all optical polarization processing, bringing shift in paradigm for classical and quantum optics applications.

* zgzheng@ecust.edu.cn
† yqlu@nju.edu.cn
# qli1@kent.edu